\documentstyle[12pt,epsfig,amsfonts]{article}
\renewcommand{\theequation}{\arabic{section}.\arabic{equation}}

\title{Radiation reaction and renormalization for a photon-like 
charged particle}

\author{\bf Yurij Yaremko\footnote{Electronic mail: yar@ph.icmp.lviv.ua}}
\date{\it Institute for Condensed Matter Physics, \\
1 Svientsitskii St., 79011 Lviv, Ukraine}
\begin{document}
\maketitle
\begin{abstract}
A renormalization scheme which relies on energy-momentum and angular
momentum balance equations is applied to the derivation of effective
equation of motion for a massless point-like charge. Unlike the massive 
case, the rates of radiated energy-momentum and angular momentum tend to 
infinity whenever the source is accelerated. The external electromagnetic 
fields which do not change the velocity of the particle admit only its 
presence within the interaction area. The effective equation of motion
is the equation on eigenvalues and eigenvectors of the electromagnetic 
tensor. The massless charges move along base line determined by the 
eigenvectors when the effective equation of motion possesses uniform 
solutions. It is interesting that the same solution arises in Rylov's model 
of magnetosphere of a rapidly rotating neutron star (pulsar).
\end{abstract}

\section{Introduction}
\setcounter{equation}{0}
In the paper \cite{CG} massless charged particles of spin one or larger are 
excluded in quantum electrodynamics by the argument that masslessness, 
Lorentz invariance, and electromagnetic coupling, are mutually incompatible. 
Roughly speaking, the interaction with an external electromagnetic field 
drastically changes incoming massless particle state, so that outgoing 
state does not describe a particle without rest mass. Further \cite{MS}
the existence of massless charges is forbidden in general by the condition 
that the energy of such particles in the electromagnetic field has no lower 
bound. In the present paper we consider the problem of reality of a massless 
charge within the realm of classical field theory.

A recent paper by Kazinski and Sharapov \cite{KS} considers the problem of 
effective equations of motions for a massless charged particle under the 
influence of its own electromagnetic field as well as an external one. 
The authors apply regularization procedure developed in their previous paper 
\cite{KLS} where the problem of radiation back reaction in classical 
electrodynamics of a point massive charge arbitrarily moving in flat 
space-time of any dimensions is studied. The 5-th order differential 
equation is derived \cite[eq.(32)]{KS} which governs the dynamics of the
photon-like charge in four dimensions. The reduction procedure is developed 
which allows to select the solutions of true physical meaning.

Since the concept of a "zero-mass interacting particle" is quite different 
in quantum and classical theories, it would be more appropriate to obtain 
the equation of motion as a limiting case of the well-known Lorentz-Dirac 
equation \cite{Dir}. (It defines the motion of point-like charge with rest 
mass $m$ under the influence of an external force as well as its own 
electromagnetic field, for a modern review see \cite{Rohr,Pois,TVW}.)
In \cite{RlJ} the motion of massive charged particles in a very strong 
electromagnetic field is studied. The guiding center approximation \cite{Fr}
is used in the Lorentz-Dirac equation. In this approximation the particle 
motion is described as a combination of forward and oscillatory motions 
(the field changes are small during a gyration period). If the gradient of 
the field potential is much larger than the rest mass of the particle, the 
strong radiation damping suppresses the particle gyration. It is shown 
\cite{RlJ} that the particle velocity is directed along one of the 
eigenvectors of the (external) electromagnetic tensor if $m\to 0$ in the 
rewritten Lorentz-Dirac equation. The equation on eigenvalues and 
eigenvectors of the electromagnetic tensor governs the motion of charges 
in the massless approximation.

According to \cite{RlJ}, the effective equation of motion for this charge 
does not contain derivatives higher than 1. This conclusion is in 
contradiction with that of Ref.\cite{KS} where the radiation back reaction 
is finite and the 5-th order differential equation determines the 
evolution of photon-like charge.

In general, the regularization procedure can be performed in two quite 
different ways: (i) one when Green's functions are used in variational equations of
motion; (ii) the other when wave solutions are substituted for field
variables in Noether conservation laws (e.g., in energy-momentum carried 
by electromagnetic field). In \cite{KS,KLS} the first way is realized 
which is a combination of some heuristic assumptions and calculations 
by methods of functional analysis. The second way is the integration 
of the Maxwell energy-momentum density over a space-like surface in 
Minkowski space. 

\begin{figure}[t]
\begin{center}
\epsfclipon
\epsfig{file=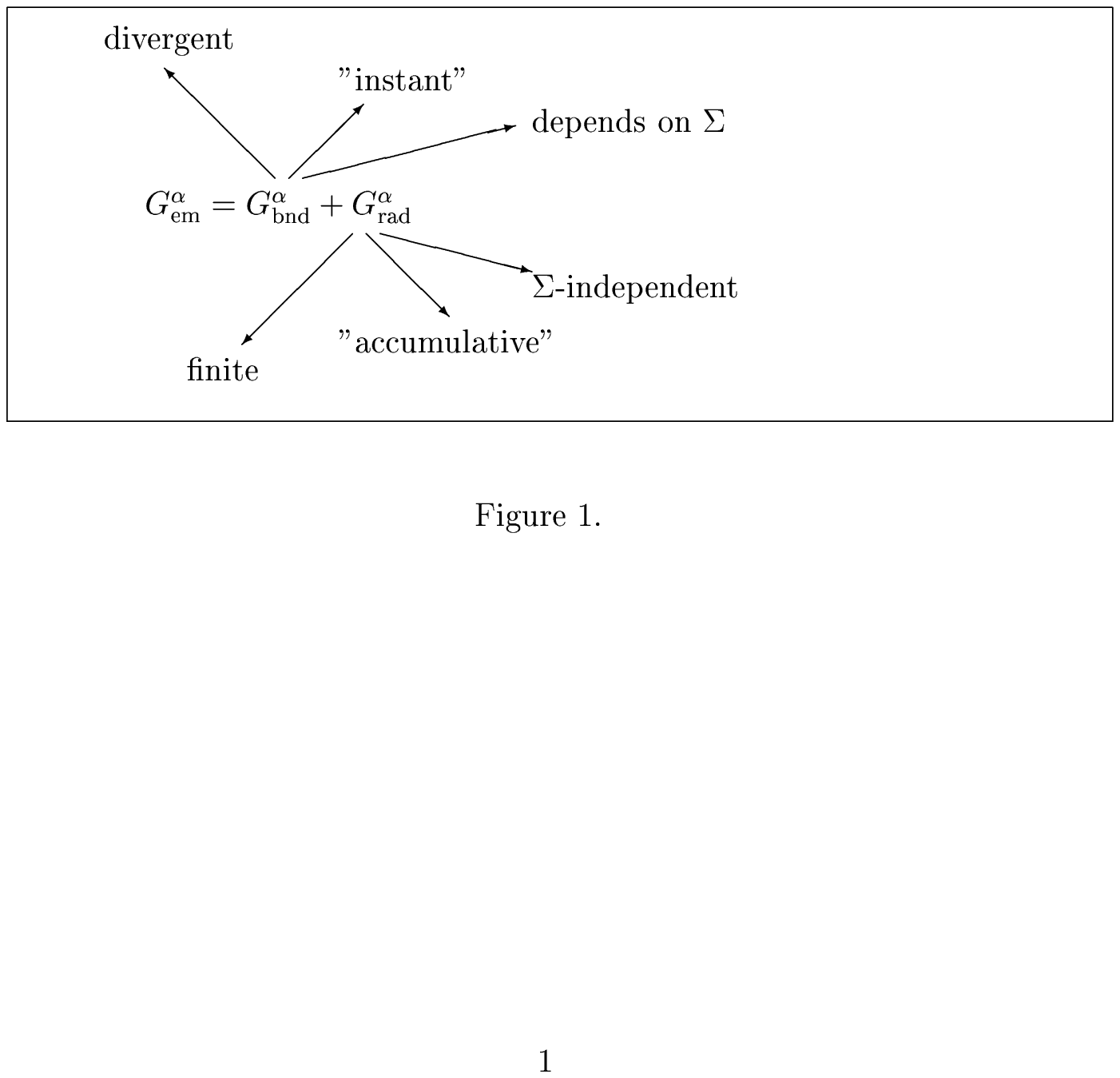,width=8cm}
\end{center}
\caption{\label{Noe}
\small The bound term $G^\alpha_{bnd}$ and the radiative term
$G^\alpha_{rad}$ constitute Noether quantity $G^\alpha_{em}$ carried by
electromagnetic field. The former diverges while the latter is finite.
Bound component depends on instant characteristics of charged particles
while the radiative one is accumulated with time. The form of the bound
term heavily depends on choosing of an integration surface $\Sigma$
while the radiative term does not depend on $\Sigma$.
}
\end{figure}

Teitelboim in \cite{Teit} classifies the terms which arise due to 
integration (see figure \ref{Noe}). Within regularization procedure the 
bound terms are coupled with energy-momentum and angular momentum of 
"bare" sources, so that already renormalized characteristics 
$G^\alpha_{part}$ of charged particles are proclaimed to be finite. 
Noether quantities which are
properly conserved become:
\begin{equation}\label{G}
G^\alpha=G^\alpha_{part}+G^\alpha_{rad}.
\end{equation}
Particle's equations of motion arise from analysis of differential 
consequences of the conserved quantities (\ref{G}). i.e. from the balance 
equations ${\dot G}^\alpha=0$.

In the present paper we apply the regularization procedure based on
Noether conservation laws to the problem of radiation reaction for a 
massless charge in response to the electromagnetic field. 

The paper is organized as follows. In Section 2 we state our notation. In 
Section 3 we discuss some peculiarities of electromagnetic field generated 
by a photon-like charge. Contrary to the massive case, the field strengths 
contain the {\it far} terms only (these scaled as $r^{-1}$ where $r$ is the 
retarded distance \cite{Rohr,Pois}). The term which is scaled as 
$r^{-2}$ exhausts corresponding {\it radiative} stress-energy tensor. 
Volume integration of the Maxwell energy-momentum tensor density gives the 
flux of radiative momentum (see Section 4). In Section 5 we present our 
main result --- the effective equations of motion for a massless charged 
particle under the influence of an external force. Since the radiation 
back reaction diverges when the particle is accelerated, the external 
device should not change its velocity. Few electromagnetic fields are 
briefly described in Appendix A which admit the photon-like charges within 
the interaction area. Finally, in Section 6 a short comment is made about 
their possible presence in the magnetosphere of a pulsar \cite{Rl8,Rl9}.

\section{General setting}
\setcounter{equation}{0}

Let ${\mathbb M}_{\,4}$ be Minkowski space with coordinates $x^\mu$ and 
metric tensor $\eta_{\mu\nu}={\rm diag}(-1,1,1,1)$. We use Heaviside-Lorentz 
system of units with the velocity of light $c=1$. Summation over repeated 
indices is understood throughout the paper; Greek indices run from $0$ to 
$3$, and Latin indices from $1$ to $3$.

We consider a massless point-like particle which carries an electric 
charge $q$ and moves on a lightlike world line $\gamma:{\mathbb R}\to 
{\mathbb M}_{\,4}$ described by functions $z^\mu(\tau)$, in which $\tau$ is 
an arbitrary parameter. A tangent vector to each point 
$z^\mu(\tau)\in\gamma$ lies on the future light cone with vertex at this 
point: 
\begin{equation}\label{I0}
{\dot z}^2=0.
\end{equation}
(We use an overdot on $z$ to indicate differentiation with respect to the 
evolution parameter $\tau$.) We let $u^\alpha(\tau)={\rm d} 
z^\alpha/{\rm d}\tau$ be the 4-velocity, and $a^\alpha(\tau)={\rm d} 
u^\alpha/{\rm d}\tau$ is the 4-acceleration. Initially we take the world 
line to be arbitrary; our main goal is to find the particle's equation of 
motion.

Following \cite{KS}, we deal with an obvious generalization of the standard 
variational principle for massive charge
\begin{equation}\label{I}
I = I_{\rm particle} + I_{\rm int} + I_{\rm field}\,,
\end{equation}
with
\begin{equation} \label{fint}
I_{\rm field}=-\frac{1}{16\pi}\int {\rm d}^4xf^{\mu\nu}f_{\mu\nu}
\qquad
I_{\rm int} = \int {\rm d}^4xA_\mu j^\mu .
\end{equation}

The particle part of variational principle should be consistent with the 
field and the interaction terms. So, if we require that the renormalized 
mass be zero, a nonzero bare mass is necessary to absorb a divergent 
self-energy. Hence the world line of the bare particle should be assumed 
time-like rather than lightlike. We may also require that the world line 
be lightlike before renormalization as well as after this procedure. To 
solve the dilemma we establish the structure of the bound and 
radiative terms (see figure \ref{Noe}) of energy-momentum and 
angular momentum carried by electromagnetic field of the photon-like 
charge. 

Having variated (\ref{fint}) with respect to potential $A_\mu$, we obtain 
the Maxwell field equations \cite[eq.(14)]{KS} 
\begin{equation}
\square A_\mu (x)=-4\pi j_\mu (x)
\end{equation}
where current density is zero everywhere, except at the particle's position 
where it is infinite
\begin{equation}\label{j}
j_\mu (x)=q\int {\rm d}\tau u_\mu (\tau)\delta[x-z(\tau)]
\end{equation}
and $\square :=\eta^{\alpha\beta}\partial_\alpha\partial_\beta$ is the 
wave operator.

The components of the momentum 4-vector carried by the electromagnetic field 
are \cite{Rohr,Pois}
\begin{equation} \label{pem}
p_{\mbox{\scriptsize em}}^\nu (\tau)=\int_{\Sigma}
{\rm d}\sigma_\mu T^{\mu\nu} 
\end{equation}
where ${\rm d}\sigma_\mu$ is the outward-directed surface element on an 
arbitrary space-like hy\-per\-sur\-fa\-ce $\Sigma$. The angular momentum 
tensor of the electromagnetic field is written as
\cite{Rohr}
\begin{equation} \label{Mem}
M_{\mbox{\scriptsize em}}^{\mu\nu}(\tau)=\int_{\Sigma}
{\rm d}\sigma_\alpha\left(x^\mu T^{\alpha\nu}-x^\nu T^{\alpha\mu}\right) 
\end{equation}
where
\begin{equation}\label{T}
T^{\mu\nu} = \frac{1}{4\pi}\left(f^{\mu\lambda}f^\nu{}_\lambda - 
1/4\eta^{\mu\nu} f^{\kappa\lambda}f_{\kappa\lambda}\right)
\end{equation}
is the electromagnetic field's stress-energy tensor. 

\section{Electromagnetic field of a photon-like charge}
\setcounter{equation}{0}
Let the past light cone with vertex at an observation point $x$ is 
punctured by the particle's world line $\gamma$ at point $z(s)$.
The retarded Green function associated with the d'Alembert operator 
$\square$ and the charge-current density (\ref{j}) is valuable 
only. The components of the Li\'enard-Wiechert potential $\hat 
A=A_\alpha{\rm d} x^\alpha$ are
\begin{equation}\label{A}
A_\alpha=q\frac{u_\alpha(s)}{r}
\end{equation}
where $r=-(R\cdot u)$ is the retarded distance \cite{Rohr,Pois}; 
$R^\mu=x^\mu-z^\mu(s)$ is 
the null vector pointing from $z(s)\in\gamma$ to $x$. The 4-potential is not 
defined at points on the ray in the direction of momentary 4-velocity $u(s)$
by reason of the isotropy condition (\ref{I0}).

Straightforward computation reveals that $\square A=0$ everywhere, 
except at the particle's position. Indeed, suppose 
that the observation point $P$ with coordinates $x$ is moved to $P'(x+\delta 
x)$. This induces a change in the intersection point $z(s)$.
The new intersection point is then $z(s+\delta s)$; points $P'(x+\delta x)$ 
and $z(s+\delta s)$ are still related by null 4-vector
$R^\mu=x^\mu+\delta x^\mu-z^\mu(s+\delta s)$. Expanding the relation $R^2=0$
to the first order in the displacements, we obtain the differentiation rule 
\begin{equation}\label{du}
\frac{\partial s}{\partial x^\alpha}=-k_\alpha \qquad 
k^\alpha=\frac{x^\alpha -z^\alpha(s)}{r}.
\end{equation}
Differentiation of the retarded distance gives
\begin{equation}\label{dr}
\frac{\partial r}{\partial x^\alpha}=-u_\alpha+ra_kk_\alpha
\end{equation}
where $a_k:=(a\cdot k)$ is the component of the acceleration $a(s)$ in the 
direction of $k$. We also need the equality
\begin{equation}\label{dk}
\frac{\partial k_\alpha}{\partial x^\beta}=r^{-1}\left(
u_\alpha k_\beta + u_\beta k_\alpha +\eta_{\alpha\beta}
\right)-a_kk_\alpha k_\beta.
\end{equation}
Finally we act on (\ref{A}) by the wave operator
\begin{equation}\label{op}
\square =\eta_{\alpha\beta}
\frac{\partial }{\partial x^\alpha}
\frac{\partial }{\partial x^\beta}
\end{equation}
Using (\ref{du}), (\ref{dr}) and (\ref{dk}), after some algebra we obtain 
zero.

Because of isotropy condition $(u\cdot u)=0$ the rules (\ref{dr}) and 
(\ref{dk}) are different from their counterparts 
\cite[eqs.(4.7),(4.9)]{Pois} for massive particle.

Unlike the massive case, the photon-like charge generates the far 
electromagnetic field $\hat f={\rm d}\hat A$:
\begin{equation}\label{f}
\hat f= q\frac{a\wedge k+a_ku\wedge k}{r}
\end{equation}
Here the dot means the scalar product of two 4-vectors and the 
symbol $\wedge$ denotes the wedge product. Because of isotropy condition 
the retarded distance vanishes on the ray in the direction of particle's 
4-velocity taken at the instant of emission. The field diverges at all the 
points of this ray with vertex at the point of emission.

To calculate the stress-energy tensor of the electromagnetic field we 
substitute the components (\ref{f}) into expression (\ref{T}). Contrary to 
the massive case \cite[eqs.(5.3)-(5.5)]{Pois}, the "photon-like" Maxwell 
energy-momentum density contains the radiative component only:
\begin{equation}\label{Trad}
4\pi T^{\alpha\beta}=\frac{q^2}{r^2}a^2k^\alpha k^\beta .
\end{equation}
Hence the divergent self-energy which is due to volume integration of the 
{\it bound} part of the electromagnetic field's stress-energy tensor 
\cite{Teit} does not arise. Unlike the massive case, the photon-like charge 
does not possess an electromagnetic "cloud" permanently attached to it. 
The renormalization procedure is not necessary because the photon-like 
source is not "dressed".

As a consequence, the Brink-Di Vecchia-Howe action term \cite[eq.(2)]{BVH}:
\begin{equation}\label{prt}
I_{\rm particle} = \frac12\int d\tau e(\tau){\dot z}^2 
\end{equation}
is consistent with the field an interaction terms (\ref{fint}). 
Variation of (\ref{prt}) with respect to Lagrange multiplier 
$e(\tau)\ne 0$ yields the isotropy condition (\ref{I0}).
The particle part (\ref{prt}) of the total action (\ref{I}) describes {\it 
already renormalized} massless charge.

The action integral (\ref{I}) being the sum of (\ref{fint}) and (\ref{prt}) 
is invariant under arbitrary time and space translations as well as space 
and mixed spacetime rotations. The Poincar\'e invariance of (\ref{I}) 
assures us, via Noether's theorem, of ten conservation laws, i.e. those 
quantities which do not change with time. 

Action integral (\ref{I}) with $I_{\rm part}$ in form of (\ref{prt}) is 
conformally invariant. This symmetry property is analyzed in Appendix B. It 
is worth noting that the conformal invariance yields conservation laws, 
which are functions of energy-momentum and angular momentum conserved 
quantities.

\section{Energy-momentum and angular momentum carried by the electromagnetic 
field }
\begin{figure}
\begin{center}
\epsfclipon
\epsfig{file=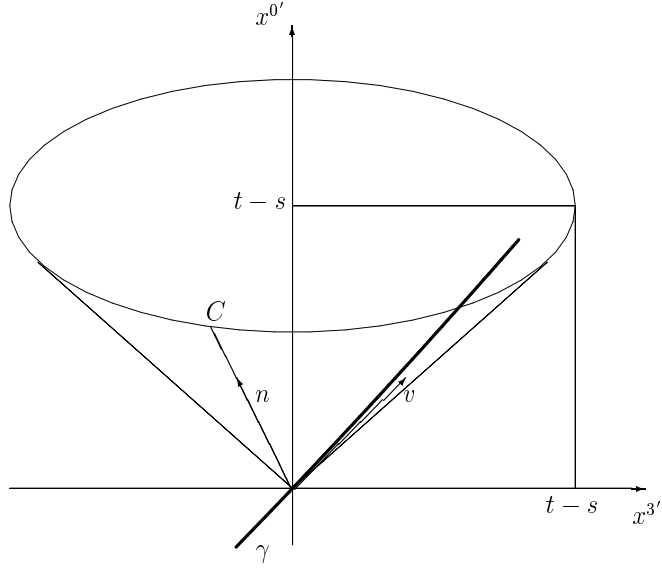,width=9cm}
\end{center}
\caption{\label{ret-d} 
In the particle's momentarily comoving frame the massless charge is placed 
at the coordinate origin; its 4-velocity is $(1,0,0,1)$. The point $C\in 
S(0,t-s)$ is linked to the coordinate origin by a null ray characterized by 
the angles $(\varphi,\vartheta)$. (The null vector 
$n=(1,\cos\varphi\sin\vartheta,\sin\varphi\sin\vartheta,\cos\vartheta)$
defines this direction.) For a given point $C$ with coordinates $x^{\alpha'}=
(t-s)n^{\alpha'}$ the retarded distance is 
$x^{0'}-x^{3'}=(t-s)(1-\cos\vartheta)$. 
}
\end{figure}

\setcounter{equation}{0}
Volume integration of the radiative energy-momentum density (\ref{Trad}) 
over a hyperplane $\Sigma_t=\{x\in {\mathbb M}_{\,4}: x^0=t\}$ gives the 
amount of radiated energy-momentum at fixed instant $t$. An appropriate 
coordinate system is a very important for the integration. We introduce the 
set of curvilinear coordinates for flat space-time ${\mathbb M}_{\,4}$ 
involving the observation time $t$ and the retarded time $s$:
\begin{equation}\label{cvc}
x^\alpha=z^\alpha(s) + (t-s)\Omega^\alpha{}_{\alpha'}n^{\alpha'}.
\end{equation}
The null vector $n:=(1,{\bf n})$ has the components 
$(1,\cos\varphi\sin\vartheta,\sin\varphi\sin\vartheta,\cos\vartheta)$;
$\vartheta$ and $\varphi$ are two polar angles. Matrix space-time components
are $\Omega_{0\mu}=\Omega_{\mu 0}=\delta_{\mu 0}$; its space components 
$\Omega_{ij}$ constitute the orthogonal matrix which rotates space axes of 
the laboratory Lorentz frame until new $z$-axis is directed along 
three-vector ${\bf v}$. (Particle's 4-velocity has the form $(1,v^i), |{\bf 
v}|=1,$ if parametrization of the world line $\gamma$ is provided by a 
disjoint union of hyperplanes $\Sigma_t$.) Orthogonal matrix
\begin{equation}\label{om}
\omega=\left(
\begin{array}{ccc}
\cos\varphi_v&-\sin\varphi_v&0\\
\sin\varphi_v&\cos\varphi_v&0\\
0 & 0 & 1
\end{array}
\right)
\left(
\begin{array}{ccc}
\cos\vartheta_v&0&\sin\vartheta_v\\
0 & 1 & 0\\
-\sin\vartheta_v&0&\cos\vartheta_v
\end{array}
\right)
\end{equation}
where 
$v^i=(\cos\varphi_v\sin\vartheta_v,\sin\varphi_v\sin\vartheta_v,\cos\vartheta_v)$ 
determines the rotation. In terms of curvilinear coordinates 
$(t,s,\vartheta,\varphi)$ the retarded distance is as follows:
\begin{equation}
r=(t-s)(1-\cos\vartheta).
\end{equation}
The situation is pictured in figure \ref{ret-d}.

So, we construct the global coordinate system centred on the world line 
of the massless particle. Minkowski space ${\mathbb M}_{\,4}$ becomes a 
disjoint union of hyperplanes $\Sigma_t=\{x\in {\mathbb M}_{\,4}: x^0=t\}$.
A surface $\Sigma_t$ is a disjoint union of spherical wave fronts
\begin{equation} \label{S}
S(z(s),t-s)=\{x\in 
{\mathbb M}_{\,4}: (x^0-s)^2=\sum_i(x^i-z^i(s))^2,x^0=t\}
\end{equation}
which are the intersections of the future light cones with 
vertices at points $z(s)\in\gamma$ and hyperplane $\Sigma_t$. The point 
$C\in S(z(s),t-s)$ is linked to the point $z(s)\in\gamma$ by a null 
ray characterized by the angles $(\varphi,\vartheta)$ specifying its 
direction on the cone. 

Now we calculate the electromagnetic field momentum
\begin{equation} \label{p_em}
p_{\rm em}^\mu = \int_{\Sigma_t}{\rm d}\sigma_0T^{0\mu}
\end{equation}
where an integration hypersurface $\Sigma_t=\{x\in {\mathbb M}_{\,4} : 
x^0=t\}$ is a surface of constant $t$. 

The surface element is given by ${\rm d}\sigma_0=\sqrt{-g}{\rm d} s 
{\rm d}\vartheta {\rm d}\phi$ where
\begin{equation} \label{se}
\sqrt{-g} = (t-s)^2\sin\vartheta (1-\cos\vartheta)
\end{equation}
is the determinant of metric tensor of Minkowski space viewed in
curvilinear coordinates (\ref{cvc}). In these coordinates the components of 
the electromagnetic field's stress-energy tensor (\ref{Trad}) have the form:
\begin{eqnarray}
4\pi T^{00}&=&q^2\frac{a^2(s)}{(t-s)^2(1-\cos\vartheta)^4}\\
4\pi T^{0i}&=&q^2\frac{a^2(s)\omega_{ii'}n^{i'}}{(t-s)^2(1-\cos\vartheta)^4}.
\end{eqnarray}
The angular integration results the radiated energy-momentum:
\begin{equation}\label{e-m}
p^0_{\rm em}=\frac{q^2}{2}I_0\int\limits_{-\infty}^t{\rm d} s a^2(s)\qquad
p^i_{\rm em}=\frac{q^2}{2}I_1\int\limits_{-\infty}^t{\rm d} s a^2(s)v^i(s)
\end{equation}
where factors $I_n$ diverge:
\begin{eqnarray}
I_0&:=&\int\limits_0^{\pi}{\rm d}\vartheta
\frac{\sin\vartheta}{(1-\cos\vartheta)^3}=
-\frac{1}{8}+\lim\limits_{\vartheta\to 0}
\frac{1}{2(1-\cos\vartheta)^2}\\
I_1&:=&\int\limits_0^{\pi}{\rm d}\vartheta
\frac{\sin\vartheta\cos\vartheta}{(1-\cos\vartheta)^3}=
\frac{3}{8}-\lim\limits_{\vartheta\to 0}
\left[\frac{1}{1-\cos\vartheta}-\frac{1}{2(1-\cos\vartheta)^2}\right].
\end{eqnarray}

Similarly, the computation of the electromagnetic field angular 
momentum which flows across the hyperplane $\Sigma_t$ gives rise to the 
divergent quantities:
\begin{eqnarray}\label{a-m1}
M_{\rm em}^{0i}&=&\frac{q^2}{2}I_1\int\limits_{-\infty}^t{\rm d} s 
a^2(s)sv^i(s)-
\frac{q^2}{2}I_0\int\limits_{-\infty}^t{\rm d} s a^2(s)z^i(s)\\
M_{\rm em}^{ij}&=&\frac{q^2}{2}I_1\int\limits_{-\infty}^t{\rm d} s 
a^2(s)\left[z^i(s)v^j(s)-z^j(s)v^i(s)\right].\label{a-m2}
\end{eqnarray}

\begin{figure}
\begin{center}
\epsfclipon
\epsfig{file=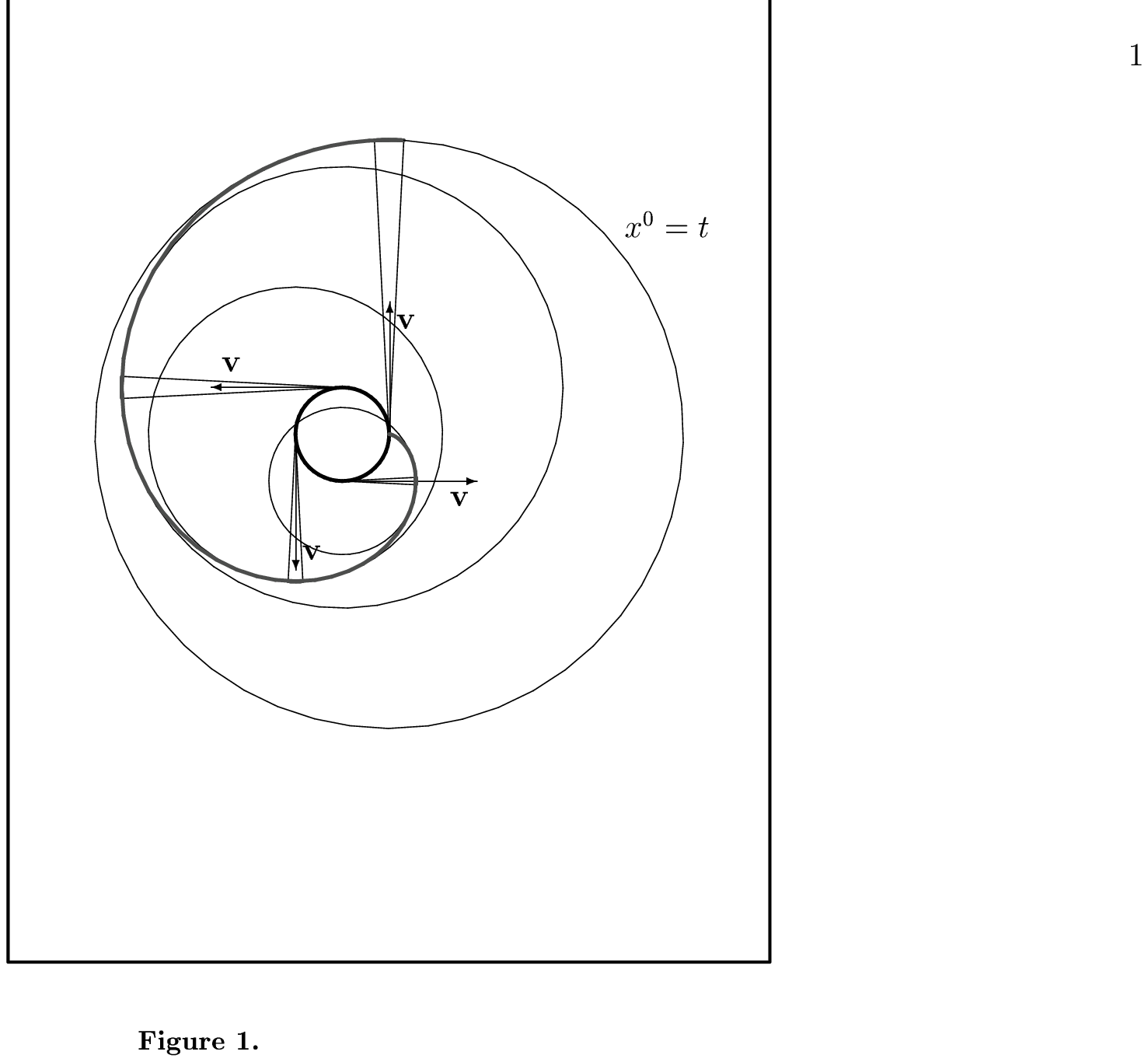,width=8cm}
\end{center}
\caption{\label{rad} 
The bold circle pictures the trajectory of a photon-like charge. The others 
are spherical wave fronts (\ref{S}) viewed in the observation hyperplane 
$\Sigma_t=\{x\in {\mathbb M}_{\,4}: x^0=t\}$. The circling photon-like 
charge radiates infinite rates of energy-momentum and angular momentum 
in the direction of its velocity ${\bf v}$ at the instant of emission.
The energy-momentum and angular momentum carried by electromagnetic field of 
accelerated charge tend to infinity on the spiral curve.
}
\end{figure}

The energy-momentum (\ref{e-m}) and the angular momentum (\ref{a-m1}) and 
(\ref{a-m2}) of electromagnetic field generated by the accelerated 
photon-like charge tend to infinity in the direction of particle's velocity 
at  the instant of emission. The divergent terms are not bound terms which 
should be absorbed by corresponding particle characteristics within the 
renormalization procedure. Indeed, they do not depend on the distance from 
the particle's world line. Secondly, the energy-momentum and the angular 
momentum accumulate with time at the observation hyperplane $\Sigma_t$ (see 
figure \ref{rad}). Hence the divergent Noether quantities cannot be 
referred to an electromagnetic "cloud" which is permanently attached to the 
charge and is carried along with it.

Changes in energy-momentum and angular momentum radiated by accelerated 
charge should be balanced by changes in already renormalized 4-momentum and 
angular momentum of the particle\footnote{If the massive charge coupled 
with electromagnetic field is considered \cite{YR,Yar1}, the balance 
equations yield the Lorentz-Dirac equation.}. But the accelerated 
photon-like charge emits infinite amounts of radiation (see 
figure \ref{rad}).
To change the velocity of the massless charge the energy is 
necessary which is too large to be observed. Does it mean that there is no 
photon-like charges within an interaction area? In the Appendix A we 
sketch several electromagnetic fields which do not change the velocity of 
the massless charge. 

\section{Massless charge within an interaction area}
\setcounter{equation}{0}
According to expression (\ref{f}), non-accelerated photon-like charge does 
not generate the electromagnetic field. The evolution of the particle beyond 
an interaction area is determined by the Brink-Di Vecchia-Howe Lagrangian 
\cite{BVH}
\begin{equation}\label{L0}
L = \frac12 e(\tau){\dot z}^2 .
\end{equation}
The particle's 4-momentum $p_{\rm part}^\mu=e(\tau){\dot z}^\mu$ does 
not change with time:
\begin{eqnarray}
{\dot p}_{\rm part}^\mu&=&{\dot e}(\tau){\dot z}^\mu\\
&=&0.\nonumber
\end{eqnarray}
Since ${\dot z}^\mu\ne 0$, the Lagrange multiplier $e$ does not depend on 
$\tau$. We deal with a photon-like particle moving in the ${\bf 
\dot{z}}$-direction with frequency
$\omega_0=e_0{\dot z}^0$, such that its energy-momentum 4-vector can be 
written $p_{\rm part}^\mu=(\omega_0,\omega_0{\bf k}), |{\bf k}|=1$.

Further in this paper we shall use a disjoint union of hyperplanes 
$\Sigma_t=\{x\in{\mathbb M}_{\,4}:x^0=t\}$ for parametrization of the 
particle world line $\gamma$. We define $v^\alpha(t)={\rm d} 
z^\alpha(t)/{\rm d} t$ as the 4-velocity; 4-acceleration $a^\alpha(t)={\rm 
d} v^\alpha(t)/{\rm d} t$ looks as $(0,{\dot v}^i)$ in this 
parametrization. Since $\gamma$ is degenerate (the condition $a^2=0$ at all 
points $z\in\gamma$ is fulfilled), the 4-acceleration vanishes.

When considering the system under the influence of an external device the 
change in particle's 4-momentum is balanced by an external force 
$F_{\rm ext}$:
\begin{eqnarray}\label{eqm}
{\dot p}_{\rm part}^\mu(t)&=&\dot{e}(t)v^\mu\\
&=&F_{\rm ext}.\nonumber
\end{eqnarray}
(The 4-vector $F_{\rm ext}$ should be orthogonal to the 4-velocity.) This 
effective equation of motion is supplemented with the condition of absence 
of radiative damping. In other words, the external device admits a massless 
charge if and only if the components of null vector of 4-velocity do not 
change with time despite the influence of the external field. The 
conclusion is similar to that of Refs.\cite{CG,MS}.

When the photon-like charged particle moves in the external electromagnetic 
field $\hat F$, the Lorentz force balances the change in its 4-momentum:
\begin{equation} \label{ef}
{\dot e}v^\mu=qF^\mu{}_\nu v^\nu. 
\end{equation} 
It is convenient to decompose $\hat F$ into an electric field ${\bf E}$ 
and a magnetic field ${\bf B}$. (\ref{ef}) is then rewritten as
\begin{eqnarray}\label{ef_a}
\dot{e}&=&q({\bf E}\cdot {\bf v})\\
\dot{e}{\bf v}&=&q{\bf E}+q[{\bf v}\times {\bf B}].\label{ef_b}
\end{eqnarray}

We have the following 4-th order algebraic equation on eigenvalues ${\dot e}$
\cite{Fr}:
\begin{equation} \label{ev}
{\dot e}^4+{\dot e}^2q^2\left({\bf B}^2-{\bf E}^2\right)-q^4({\bf 
B}\cdot{\bf E})^2=0.
\end{equation} 
In general, it possesses two real solutions \cite{RlJ,Fr}
\begin{equation}\label{eig}
{\dot e}_\pm=\pm q\sqrt{\left({\bf E}^2-{\bf B}^2+\mu\right)/2}\qquad
\mu=\sqrt{\left({\bf B}^2-{\bf E}^2\right)^2+4({\bf B}\cdot{\bf E})^2} .
\end{equation}
The field admits a photon-like charge if and only if corresponding 
eigenvectors
\begin{equation}\label{sgen}
{\bf v}_\pm=\frac{[{\bf E}\times{\bf B}]\pm\left(\lambda{\bf 
E}+\kappa\nu{\bf B}\right)}{\sigma}\qquad
\kappa={\rm sgn}[({\bf B}\cdot{\bf E})]
\end{equation}
are of constant values. Here
\begin{equation}
\lambda=\sqrt{\left({\bf E}^2-{\bf B}^2+\mu\right)/2}\quad 
\nu=\sqrt{\left({\bf B}^2-{\bf E}^2+\mu\right)/2}\quad 
\sigma=\left({\bf E}^2+{\bf B}^2+\mu\right)/2.
\end{equation}

The expression (\ref{sgen}) is obtained 
in \cite[eq.(2.3)]{Rl9} where the model of magnetosphere of a rapidly 
rotating neutron star (pulsar) is elaborated. It defines the velocity of the 
massless charged particles which constitute the so-called "dynamical phase" 
of the gas of ultrarelativistic electrons and positrons moving in a very 
strong electromagnetic field of the pulsar. In Rylov's model \cite{Rl8,Rl9} 
the massless charges as a limiting case of massive ones are considered. The 
reason is that the gradient of star's potential is much larger than 
the particle's rest energy $m_ec^2$.

\section{Conclusions}
Our consideration is founded on the Maxwell equations with point-like 
source which governs the propagation of the electromagnetic field produced 
by a photon-like charge. Unlike the massive case, it generates the {\it 
far} electromagnetic field which does not yield to divergent Coulomb-like 
self-energy. Hence the world line is null before renormalization
as well as after this procedure. We choose Brink-Di Vecchia-Howe action 
\cite{BVH} for a bare particle moving on the world line which is proclaimed 
then to be lightlike.

A surprising feature of the study of the radiation back reaction in dynamics 
of the photon-like charge is that the Larmor term diverges whenever the 
charge is accelerated. Since the emitted radiation detaches itself from the 
charge and leads an independent existence, it cannot be absorbed within a 
renormalization procedure.

Inspection of the energy-momentum and angular momentum carried by the 
ele\-ctro\-mag\-ne\-tic field of a photon-like charge reveals the reason why 
it is not yet detected (if it exists). Noninteracting massless charges do 
manifest themselves in no way. Any external electromagnetic field (including 
that generated by a detecting device) will attempt to change the velocity of 
the charged particle. Whenever the effort will be successful, the radiation 
reaction will increase extremely. In general, this circumstance forbids the 
presence of the photon-like charges within the interaction area. 

Nevertheless, there exists the electromagnetic fields which do not change 
the velocities of the massless charged particles. For example, 
superposition of plane waves propagating along some base line admits the 
massless charges moving analogously. (But any disturbance annuls such a 
"loyalty".) It is worth noting that the quantum mechanical results 
\cite{CG,MS} are in favour the conception that the external field 
distinguishes the directions of non-accelerating motions of photon-like 
charges (if they exist). 

To survive photon-like charges need an extremely strong field of specific 
con\-fi\-gu\-ra\-ti\-on, as that of the rotating neutron star (pulsar). In 
\cite{Rl8,Rl9} the model of the pulsar magnetosphere is elaborated. It 
involves the so-called {\it dynamical phase} which consists of the massless 
charged particles moving with speed of light along some base line determined 
by the electromagnetic field of the star\footnote{The massless approximation is meant where the 
gradient of star's potential is much larger than electron's rest energy.}. 
It is worth noting that the expression for the particles' velocity 
\cite[eq.(2.2)]{Rl9} coincides with the solution (\ref{sgen}) of the 
"massless" equations of motion derived in the present paper.

Equation (\ref{ef}) on eigenvalues and eigenvectors of the electromagnetic 
tensor governs the motion of charges in zero-mass approximation. This 
conclusion is in contradiction with that of Ref.\cite{KS} where the 
radiation back reaction is finite and the 5-th order differential equation 
determines the evolution of photon-like charge. The reason is that 
regularization approach to the radiation back reaction (smoothing the 
behaviour of the Lorentz force in the immediate vicinity of the particle's 
world line), employed by Kazinski and Sharapov, is not valid in the case of 
the photon-like charged particle and its field. Indeed, the field diverges 
not only at point of world line but at all points of the ray in the 
direction of particle's 4-velocity taken at the instant of emission.
The ray singularity is stronger that $\delta$-like singularity of Green's 
function involved in \cite{KS} in the self-force expression. Hence 
integration over world line does not yield a finite part of the self force.

Conformal invariance of our particle plus (external) field system reinforce 
our conviction that the back-reaction force vanishes. Indeed, the 
appropriate renormalization procedure should preserve this symmetry property 
while the Brink-Di Vecchia-Howe action term does not contain a parameter to 
be renormalized. Therefore, the photon-like charge must not radiate.

\section*{Acknowledgments}

The author would like to thank B.P.Kosyakov, Yu.A.Rylov, V.Tre\-tyak,  and 
A.Du\-vi\-ryak for helpful discussions and critical comments.

\subsection*{\large\bf Appendix A Photon-like charges within 
the interaction area}
\renewcommand{\theequation}{A.\arabic{equation}} 
\setcounter{equation}{0}
{\it Plane wave}

In case of a plane wave moving in the positive $z-$direction, the electric 
and magnetic fields are related to each other as follows:
\begin{equation}
E_x=B_y\qquad E_y=-B_x\qquad E_z=B_z=0.
\end{equation}
Since ${\bf B}^2-{\bf E}^2$ as well as $({\bf B}\cdot{\bf E})$ vanish, the 
eigenvalues' equation (\ref{ev}) get simplified:
\begin{equation} \label{evw}
{\dot e}^4=0.
\end{equation} 
The eigenvector corresponding to the fourthly degenerate eigenvalue 
${\dot e}=0$ is defined by \cite{RlJ}
\begin{equation} 
{\bf v}=\frac{[{\bf E}\times{\bf B}]}{{\bf B}^2}={\bf n}_z.
\end{equation} 
Hence the plane wave admits massless charges moving along $z$-line in the 
positive direction. Their frequencies do not change with time.

{\it Uniform static electric field}

When ${\bf B}=0$ the equation (\ref{ev}) becomes
\begin{equation} \label{evE}
{\dot e}^4-{\dot e}^2q^2{\bf E}^2=0.
\end{equation} 
If ${\dot e}=0$ then ${\bf E}$ vanishes (see equations (\ref{ef_a}) and 
(\ref{ef_b}) and the charge's velocity is completely arbitrary (free 
particle).

The others are ${\dot e}_+=q|{\bf E}|$ and ${\dot e}_-=-q|{\bf E}|$. The 
photon-like charge moves in the direction ${\bf n}_E={\bf E}/|{\bf E}|$ or 
in the opposite one. Its 4-momentum
\begin{equation}
p_{\rm part}^0=\omega_0\pm q|{\bf E}|t
\qquad p_{\rm part}^i=\pm\omega_0n_E^i +qE^it
\end{equation} 
heavily depends on the time.

{\it Constant magnetic field}

When considering the magnetic field of constant value, the equation 
(\ref{ev}) looks as follows
\begin{equation} \label{evB}
{\dot e}^4+{\dot e}^2q^2{\bf B}^2=0.
\end{equation} 
The only real solution is the doubly degenerate trivial eigenvalue. Since 
$[{\bf v}\times{\bf B}]=0$, massless charges move along the base line 
determined by ${\bf B}$. The magnetic field does not change their 4-momenta.

{\it Orthogonal constant electric and magnetic fields}

Since $({\bf B}\cdot{\bf E})=0$, the basic equation (\ref{ev}) becomes
\begin{equation} \label{evEB}
{\dot e}^4+{\dot e}^2q^2\left({\bf B}^2-{\bf E}^2\right)=0.
\end{equation} 
Two unit three-vectors ${\bf v}$ which satisfy the {\it force-free 
approximation} \cite[eq.(1.5)]{Rl9} 
\begin{equation} \label{EB}
{\bf E}+[{\bf v}\times {\bf B}]=0
\end{equation} 
correspond to the doubly degenerate eigenvalue ${\dot e}=0$. 
After some algebra we arrive at
\begin{equation} \label{uEB}
{\bf v}_\pm=\frac{[{\bf E}\times{\bf B}]\pm 
{\bf B}\sqrt{{\bf B}^2-{\bf E}^2}}{{\bf B}^2}.
\end{equation} 

The condition $({\bf B}\cdot{\bf E})=0$ supplemented with the inequality 
$|{\bf E}|<|{\bf B}|$ defines the {\it capture surface} in Rylov's model of 
pulsar magnetosphere \cite{Rl9}. Massive particles (electrons and positrons) 
are captured in the immediate vicinity of this surface. Their kinetic 
energies vanish; they constitute the so-called {\it statical phase}. 
Nevertheless, the photon-like charges move across the capture surface with 
the velocity (\ref{uEB}). The region of pulsar magnetosphere where there 
are the dynamical phase and the statical phase is called {\it leaky capture 
region} in Refs.\cite{Rl8,Rl9}.

If $|{\bf E}|>|{\bf B}|$, then two eigenvalues ${\dot e}=\pm\sqrt{{\bf 
E}^2-{\bf B}^2}$ are valuable. Corresponding eigenvectors are
\begin{equation} \label{EBu}
{\bf v}_\pm=\frac{[{\bf E}\times{\bf B}]\pm 
{\bf E}\sqrt{{\bf E}^2-{\bf B}^2}}{{\bf E}^2}.
\end{equation} 
Having integrated ${\dot e}$ over time variable, we are sure that the 
4-momenta of photon-like charges moving with the velocities (\ref{EBu})
depend on time $t$:
\begin{equation}
p_{\rm part}^\mu=\left[\omega_0\pm\sqrt{{\bf E}^2-{\bf B}^2}t\right]{\bf 
v}_\pm .
\end{equation} 

\subsection*{\large\bf Appendix B. Conformal invariance of the effective 
equation of motion}
\setcounter{equation}{0}
\renewcommand{\theequation}{B.\arabic{equation}} 
According to \cite{FN,FRW}, conformal group 
${\cal C}(1,3)$ consists of Poincar\'e transformations (time and space 
translations, space and mixed space-time rotations), dilatations
\begin{equation}\label{dil}
x^{'\mu}=e^\theta x^\mu
\end{equation}
and conformal transformations
\begin{equation}\label{ct}
x^{'\mu}=\frac{x^\mu-b^\mu (x\cdot x)}{D},\qquad D=1-2(x\cdot b) +
(x\cdot x)(b\cdot b).
\end{equation}
(The scalar $\theta$ and 4-vector $b$ are group parameters.)

The components of electromagnetic field are transformed as follows:
\begin{equation}\label{ftr}
F_{\alpha\beta}=e^{2\theta}F'_{\alpha\beta}, \qquad 
F_{\alpha\beta} = F'_{\mu\nu}\Omega^\mu{}_\alpha \Omega^\nu{}_\beta 
\end{equation} 
where matrix
\begin{equation}
\Omega^\mu{}_\alpha:=\frac{\partial x^{'\mu}}{\partial x^\alpha}=
D^{-1}\lambda^\mu{}_\beta(x'')\lambda^\beta{}_\alpha(x),\quad
x''=\frac{x}{(x\cdot x)}-b,\quad 
\lambda^\beta{}_\alpha(x)=\delta^\beta{}_\alpha-\frac{2x^\beta 
x_\alpha}{(x\cdot x)}
\end{equation} 
satisfies the condition
\begin{equation}
\eta_{\mu\nu}\Omega^\mu{}_\alpha\Omega^\nu{}_\beta=D^{-2}\eta_{\alpha\beta} .
\end{equation} 
Since
\begin{equation}
{\dot z}^{'\mu} =e^\theta {\dot z}^\mu
\qquad {\dot z}^{'\mu} = \Omega^\mu{}_\alpha {\dot z}^\alpha
\end{equation} 
the Lagrange multiplier $e(\tau)$ involved in 
the Brink-Di Vecchia-Howe action term (\ref{prt}) transforms as
\begin{equation}
e(\tau)=e^{2\theta}e'(\tau),\qquad e(\tau)=D^{-2}e'(\tau).
\end{equation} 
Direct calculation shows, that the effective equation of motion (\ref{ef})
is invariant with respect to dilatation (\ref{dil}) and conformal 
transformation (\ref{ct}).


\end{document}